\documentclass [iop]{emulateapj}
\usepackage{graphicx}

\begin{document}

\title{Measuring Lensing Magnification of Quasars by Large Scale Structure using the Variability-Luminosity Relation}

\author{Anne H. Bauer\altaffilmark{1,2}, Stella Seitz\altaffilmark{2,3}, Jonathan Jerke\altaffilmark{4}, Richard Scalzo\altaffilmark{4,5}, David Rabinowitz\altaffilmark{4}, Nancy Ellman\altaffilmark{4}, Charles Baltay\altaffilmark{4}}
\email{bauer@ieec.uab.es}
\altaffiltext{1}{Institut de Ci\`encies de l'Espai, CSIC/IEEC, F. de Ci\`{e}ncies, Torre C5 par-2, Barcelona 08193, Spain}
\altaffiltext{2}{Universit\"{a}ts-Sternwarte M\"{u}nchen, Scheinerstr. 1, D-81679 M\"{u}nchen, Germany}
\altaffiltext{3}{Max-Planck-Institut f\"{u}r Extraterrestrische Physik, Giessenbachstr., D-85748 Garching bei M\"{u}nchen, Germany}
\altaffiltext{4}{Yale University, Department of Physics, P.O. Box 208120, New Haven, CT 06520-8120, USA}
\altaffiltext{5}{Mount Stromlo Observatory, The Australian National University, Cotter Road, Weston ACT 2611, Australia}

\begin{abstract}

We introduce a technique to measure gravitational lensing magnification 
using the variability of type I quasars.  Quasars' variability amplitudes 
and luminosities are tightly correlated, on average.  Magnification due 
to gravitational lensing increases the quasars' apparent luminosity, while 
leaving the variability amplitude 
unchanged.  Therefore, the mean magnification 
of an ensemble of quasars can be measured through the mean shift in the 
variability-luminosity relation.  As a proof of principle, we use this 
technique to measure 
the magnification of quasars spectroscopically identified in the 
Sloan Digital Sky Survey, due to gravitational lensing by galaxy clusters 
in the SDSS MaxBCG catalog.  
The Palomar-QUEST Variability Survey, reduced using the DeepSky pipeline, 
provides variability data for the sources.  
We measure the average quasar magnification as a function of scaled 
distance ($r/R_{200}$) from the nearest cluster;  our measurements are 
consistent with expectations assuming NFW cluster profiles, particularly 
after accounting for the known uncertainty in the clusters' centers.  
Variability-based lensing 
measurements are a valuable complement to shape-based 
techniques because their systematic errors are very 
different, and also because the variability measurements 
are amenable to photometric errors of a few percent and to depths seen in 
current wide-field surveys.  Given the data volume expected from current 
and upcoming surveys, 
this new technique has the potential to be competitive with weak lensing 
shear measurements of large scale structure.

\end{abstract}


\section{Introduction}

Gravitational lensing of an object by mass along the line of sight produces 
two measurable effects: shape distortion and magnification of the source.  
It is usually impossible to measure the level of magnification 
since one typically does not know the intrinsic flux of the source.  
Because of this 
fundamental problem, the field of weak lensing has focused on shape 
measurements of sources since 
one can usually assume that galaxies have no preferred orientation.  
(For a review of gravitational lensing theory and techniques, see 
\citealt{schneider04}.)
However, if the intrinsic luminosity of sources can be estimated, 
then lensing magnification can be used to study large scale structure.  

Recently, large scale sky surveys have become efficient and prolific 
enough such that variability studies can be conducted on large samples 
of rare objects.  For example, ensembles of type I (broad-line) 
quasars have been 
studied for the purpose of probing the physical causes of their common 
and dramatic variability (e.g., \citealt{vandenberk04}, \citealt{devries05}, 
\citealt{wilhite08}, \citealt{bauer09a}, \citealt{kelly09}, \citealt{macleod10}, 
\citealt{meusinger11}).  
In these works, the mean ensemble variability amplitude has been seen to correlate 
with a number of properties of the objects: e.g., time scale of the variability, 
wavelength of observation, redshift, mass, and luminosity of the systems.  
In particular, a strong anti-correlation 
has repeatedly been observed between variability amplitude and quasar luminosity, 
particularly when comparing quasars of similar mass.

Given a measurement of a quasar's variability amplitude, 
one can estimate its luminosity using this empirical relation.  
If the quasar were gravitationally lensed by intervening mass, 
the magnification effect would alter the observed luminosity.  However,
the fractional variability would remain unaffected, as the magnification, 
which is multiplicative on the luminosity, cancels.  Therefore, 
magnification due to gravitational lensing will shift the quasar's position 
in variability-luminosity space.  Quantification of this shift 
constitutes a measurement of the quasar's lensing magnification. 
Because the variability-luminosity relation has a large intrinsic variance, 
the estimate of a single quasar's magnification is not precise.  However, 
because the relation is well-determined in the mean, an ensemble of quasars can 
yield a significant measurement of the lensing effect.

In this paper we 
take advantage of the variability-luminosity relation seen in type I 
quasars to measure their magnification due to galaxy clusters 
along the line of sight.  
We use the Palomar-QUEST Variability Survey to measure the lensing 
magnifications of 3573 quasars, and compare the signal to that 
expected from gravitational lensing due to the 13,823 galaxy clusters in 
the Sloan Digital Sky Survey MaxBCG catalog (\citealt{maxbcg}).  
We measure an average cluster profile shape that is consistent with previous 
studies of the MaxBCG cluster catalog, showing that this method of 
measuring lensing magnification is 
indeed effective.

\section{AGN Structure and Quasar Optical Variability}

An active galactic nucleus (AGN) is typically described as having 
several prominent structural components.  The central black hole is 
fed by an accretion disk, which radiates a continuum of flux 
primarily in the optical and UV bands.  Offset from the 
plane of the disk 
exist clumps of gas which absorb the disk's flux and 
reradiate it as broad emission lines (hence the name broad line 
region (BLR)).  Emission from the disk and the BLR constitute 
the majority of the optical flux observed in type I quasars, 
which are the highly luminous AGN for which the observer has a 
direct line of sight to these two structures.  (Type I 
quasars will hereafter be referred to simply as quasars.)  
The detailed 
geometry of the disk and BLR are not well understood;  they 
may be part of one continuous structure (e.g. \citealt{nicastro03}, 
\citealt{elvis00}, \citealt{lovegrove10}), but they traditionally 
are described as being distinct.
A minority of quasars also show evidence 
of radio emission from a jet.  The fraction of quasars 
that have strong radio emission (i.e. are radio loud) is 
uncertain;  estimates range 
from roughly 5 to 25\% \citep{zamfir08}.  
Jet emission can account for 
a portion of a quasar's optical flux, but the fraction of optical 
flux that is jet-based depends strongly on the properties and 
orientation of the particular object.

Radio-loud quasars are on average more 
optically variable than radio-quiet quasars;  in particular, their 
short-timescale variability is enhanced (e.g., \citealt{kelly09}).  
This distinction implies that 
the jet contributes to the optical variability in these systems, most 
likely due to the dissipation of shocks.

Not all quasar variability is due to jet physics;  quasars with 
no evidence of a jet also exhibit significant flux variations.  
\cite{sesar07}, using the Sloan Digital Sky Survey (SDSS), found that 
at least 90\% of quasars are variable by at least 0.03 magnitudes 
on timescales of up to several years.  The variability of 
individual nearby radio-quiet quasars (and their lower-luminosity 
counterparts, Seyfert I galaxies) have been studied in detail using the 
technique of reverberation mapping (e.g. \citealt{blandford82}, 
\citealt{bentz09}):  spectra of the AGN are regularly obtained, 
and the variability of the spectral features is measured as a function 
of time.  Such monitoring has revealed that the disk continuum 
fluctuations are followed by fluctuations in the broad emission lines 
that echo the continuum variation.  The time lag between the continuum 
and broad line variability constitutes a measurement of the BLR size, 
which is typically on the order of light-days.  These measurements 
show that the primary source of variability in these objects is 
inside the BLR, presumably from the accretion disk and instabilities 
therein.

Accretion disk instabilities have long been a prime suspect as the 
main source of quasar variability.  \cite{kawaguchi98} calculated 
the expected relationship between variability amplitude and timescale 
based on a model of disk instabilities, as well as a 
relation expected from starbursts 
in the quasar host galaxy.  \cite{hawkins02} extended this work 
to include a prediction based on microlensing of the quasars.  A 
number of studies of ensemble quasar variability have measured the 
average variability amplitude versus timescale relation, which best agrees 
with the accretion disk instability calculation 
(\citealt{collier01}, \citealt{vandenberk04}, \citealt{devries05}, 
\citealt{wilhite08}, \citealt{adam}, \citealt{bauer09a}, \citealt{meusinger11}).  
However, this type of analysis would benefit from updated predictions 
using more complex models.

Disk instability models invoke relatively small, short-lived flares from 
numerous instabilities that combine to produce a 
stochastic light curve.  
These fluctuations could be prompted by changes 
in, for example, the accretion rate (e.g., \citealt{li08}) or in a 
magnetic field (\citealt{hirose09}).  
Stochastic light curves have been succesfully used to describe 
optical observations of large samples of quasars (\citealt{kelly09}, 
\citealt{kozlowski10}, \citealt{macleod10}).  These works model the 
quasar light curves as a damped random walk, using only three parameters: 
the typical amplitude of the short-timescale variability, the damping 
timescale, and the average flux of the light curve.  
Such light curves, 
generated from a constant component plus similar flares with 
random start time, follow the variability-luminosity relation 
\begin{equation}
\frac{\Delta L}{\overline{L}} \propto \overline{L}^{-\delta}, 
\label{poissonian_equ}
\end{equation}
where $L$ is the object's luminosity and $\delta$ depends on the 
details of the flares (\citealt{cidfernandes96}, \citealt{cidfernandes00}).  
Qualitatively, the anti-correlation simply notes that, given similar 
flare properties between two quasars, the fractional variability 
$\Delta L / \overline{L}$ will be smaller in the quasar whose 
mean luminosity $\overline{L}$ is larger.  
This prediction indeed describes the variability-luminosity 
anti-correlation seen in quasar ensembles (\citealt{vandenberk04}, 
\citealt{bauer09a}).

Alternatively, it is possible that much of the observed relationship 
between variability 
and luminosity is a secondary effect of a correlation between, 
for example, variability and the quasar's Eddington ratio.  
The Eddington ratio of a quasar describes its accretion rate, and 
can be estimated as a ratio between the quasar's luminosity and its 
mass.  \cite{ai10} find the correlation between Eddington ratio 
and variability to be more significant than that 
between luminosity and variability, using a sample of several hundred 
SDSS Seyfert I galaxies.  The authors note the positive correlation 
between the accretion rate and the disk radius which emits at a 
given wavelength.  Therefore an anti-correlation between variability 
amplitude and Eddington ratio implies that the outer regions of the 
disk show less variability than the inner radii (regardless of the 
physical mechanism behind the variability).  This position-dependence 
qualitatively agrees with the observed anti-correlation between 
variability and wavelength (\citealt{wilhite05}), if 
the hotter, bluer, inner areas of the disk vary more strongly than 
the redder, outer regions.  The variability amplitude may be 
anti-correlated with the quasar's luminosity via such a positional 
dependence, and therefore through the definition of the 
Eddington ratio as related to the luminosity.  
As described in section \ref{mu_meas_section}, 
we normalize the quasar variability measurements according 
to the objects' physical properties, including black hole mass.  
This procedure effectively allows us to measure how the variability 
amplitude scales with luminosity, for quasars of similar mass.  Holding 
the mass constant in this way allows the variability-luminosity 
relation to contain the same information as the correlation between the 
variability and the Eddington ratio.

\section{Measuring $\mu$} \label{mu_meas_section}

The average amplitude of quasar variability has been seen 
to depend on several factors: time lag between measurements, 
luminosity of the quasar 
(\citealt{vandenberk04}, \citealt{devries05}, \citealt{wilhite08}, 
\citealt{bauer09a}, \citealt{macleod10}, \citealt{meusinger11}), 
wavelength of observation (\citealt{vandenberk04}, \citealt{devries05}, 
\citealt{meusinger11}), 
and  black hole mass (\citealt{wilhite08}, \citealt{bauer09a}).  
The dependence of variability amplitude on redshift is less obvious;  
\cite{vandenberk04} measured a slight increase in the variability 
with redshift, while \cite{devries05} measured a slight decrease.  
More recently, \cite{macleod10} and \cite{meusinger11} have measured no 
significant dependence of variability amplitude on redshift.  

In practice, the variability-luminosity trend measured is of the form:
\begin{equation}
\log(V) = C - \alpha \times \log(L_{\mathrm{meas}}).
\label{v_vs_l_equation}
\end{equation}
A linear relation has indeed been observed (\citealt{vandenberk04}, 
\citealt{bauer09a}), although there are conflicting results for the 
value of the power-law slope $\alpha$, perhaps due to 
selection effects.   
If faint quasars are included in the analysis for which one cannot 
observe the full extent of the variability, the measured slope will 
become artificially shallow;  this effect most clearly manifests itself 
as a flattening in the relation at the lowest observable luminosities, 
as is illustrated in figure 5 of \cite{bauer09a}.  The low-luminosity 
limit of the binning scheme in this work is chosen to exclude this 
regime from the data set.  The value of the constant $C$ 
depends on the details of the normalization of the data, as described 
below.  

When studying how the variability 
of a large quasar sample depends on one of the quasars' 
properties, we must treat the parameters as independently as possible.  
To this end, we use a method 
introduced by \cite{vandenberk04} and adopted in \cite{bauer09a}.  
Four basic quantities are known for all of the quasars in our sample:  
time lag between measurements $\tau$, luminosity $L$, estimated black 
hole mass $M$, and redshift $z$.  There are known correlations 
between all of these parameters, due to physical relationships or 
artificial effects such as detection biases in flux-limited surveys.  
To avoid these complications and study only the dependence of variability 
on luminosity, we would like to identify a set of quasars with identical 
properties except for their luminosity, and then examine how the 
variability differs between them.  To approximate this procedure, we 
have split each parameter's range into bins: 8 bins in $\tau$, 6 bins 
in $M$, 6 bins in $z$, and 4 bins in $L$.  The bin limits are given 
in table \ref{bin_limits};  quasars with properties outside the given 
ranges are not used in the analysis.

\begin{table}
\begin{center}
\begin{tabular}{|llll|}
\hline
$\tau$ & $z$ & $M$ & $L$ \\
\hline
1 & 0.40 & 1 & 30.85 \\
5 & 0.80 & 4 & 31.05 \\
10 & 1.10 & 8 & 31.20 \\
20 & 1.40 & 12 & 31.40 \\
60 & 1.65 & 20 & 31.50 \\
130 & 1.90 & 30 &  \\
160 & 2.20& 75 & \\
220 & & & \\
400 & & & \\
\hline
\end{tabular}
\end{center}
\caption{Bin limits used in determining the normalization constants.  Measurements with values outside the limits are not considered.  Units of time lag $\tau$: 
days; black hole mass $M$: $10^{8} \times M_{\odot}$, luminosity $L$: $\frac{\mathrm{erg}}{\mathrm{s} \cdot \mathrm{Hz}}$.}
\label{bin_limits}
\end{table}

To measure quasar variability we use a quantity similar to that of the structure 
function.  We define: 
\begin{equation}
V = \sqrt{(\Delta m)^{2} - \sigma^{2}},
\label{v_equation}
\end{equation}
where $\Delta m$ is the magnitude difference between two independent 
observations of an object, and $\sigma$ is the error on those measurements.  
This is similar to the structure function as used in \cite{vandenberk04} and 
\cite{bauer09a}; however, here instead of being an ensemble measurement, 
one $V$ is measured for each pair of magnitude 
measurements of a quasar.  Four measurements of a quasar will 
yield six $\Delta m$ measurements, and therefore six different $V$ measurements 
for the single quasar.  As $V$ is imaginary when $\Delta m$ is less than the 
measurement error $\sigma$, we only use data which show significant ($>1 \sigma$) 
variability.  This cut on the data is described further in section \ref{datacuts_section}.  

For each multi-dimensional bin, a mean variability amplitude $\overline{V}$ is 
determined by taking the mean of all $V$ values measured for that bin.  Then, 
holding constant the indices for time lag, mass, redshift, and wavelength, 
one can compare the $\overline{V}$ values across the 4 $L$ bins.  This procedure yields 
$8 \times 6 \times 6 = 288$ possible 4-point plots of mean variability amplitude 
$\overline{V}$ versus luminosity, or 1152 possible $\overline{V}$ values.  Most of the 
multi-dimensional bins are not well populated by the quasar sample (for 
example, high-redshift low-luminosity bins).  In fact, we obtain 403 bins 
with at least 50 measurement pairs, which is the minimum we require in 
order to adequately determine $\overline{V}$.  To examine the overall behavior 
of $\overline{V}$ 
with respect to $L$ one can normalize the 4-point measured trends together 
and average the resulting normalized data in each $L$ bin to find a 
simple, meaningful result of how the variability scales with the quasar 
luminosity.  The normalization consists of an additive constant in log($\overline{V}$), 
i.e. each 4-point measured trend has its own constant $C$ as defined in 
equation \ref{v_vs_l_equation}.  
The $\tau$, $M$, and $z$ multi-dimensional bin that has the best 
statistics is chosen to be the standard, and the log($\overline{V}$) versus log($L$) 
trends from all other $\tau$, $M$, and $z$ bins are normalized to 
that standard using one constant offset per $\tau$, $M$, $z$ combination.  
The constant is determined by minimizing the chi square difference between 
the $\overline{V}$ values from the two datasets in the same $L$ bin, for the $L$ bins 
where there exist data from both sets.  For a visual representation of the 
normalization procedure, see figure 4 in \cite{bauer09a}.  After averaging 
the normalized data, we are left with one $V_{\mathrm{norm}}$ versus $L$ trend with 
arbitrary $y$ axis normalization but meaningful slope.  This normalization 
technique has been shown to give results for variability amplitude versus time 
lag $\tau$ that are 
consistent with independent measurements (see table 4 in \cite{bauer09a}.  
In this way, we study how the 
variability scales with luminosity, comparing only objects with 
similar values of the other parameters.  
After the normalization, deviations from the mean $V_{\mathrm{norm}}-L$ relation will 
not be caused by known, but lensing-independent, correlations such 
as that between variability and time lag.  
Using the normalization constants calculated in this way, each measured 
variability amplitude $V$ is normalized 
according to its $\tau$, $M$, and $z$ bin;  
the resulting $V_{\mathrm{norm}}$ then can be used to estimate the quasar's 
lensing magnification, as detailed below.

We note that, for data taken in a single pass-band, the redshift of the 
quasar is degenerate with the rest-frame wavelength observed:  the 
effective wavelength of the filter in the quasar's rest frame scales as 
$(1+z)$.  Because rest-frame wavelength of observation scales 
monotonically with redshift, these two 
variables are simultaneously accounted for in our normalization versus redshift.

When fitting for $\alpha$ and $C$ in equation \ref{v_vs_l_equation} , we 
find the linear fit that best describes the whole quasar sample.  
On average, the measured luminosity $L_{\mathrm{meas}}$ of quasars in the sample 
is equal to $\overline{\mu} L_{\mathrm{intrinsic}}$, where $\overline{\mu}$ is 
a magnification value representative of the entire sample.  We can therefore 
rewrite equation \ref{v_vs_l_equation} as 
\begin{equation}
\log(V_{\mathrm{norm}}) = C - \alpha \times \log(\overline{\mu} L_{\mathrm{intrinsic}})
\label{v_vs_l_equation2}
\end{equation}
where we have also explicitly noted the fact that the $V$ measurements 
are normalized. 

Once we have determined the unique constants $\alpha$ and $C$ for the whole 
data set, we can use a measured and normalized $V_{\mathrm{norm}}$ of a quasar to 
constrain $\overline{\mu} L_{\mathrm{intrinsic}}$ since equation 
\ref{v_vs_l_equation2} is equivalent to 
\begin{equation}
\overline{\mu} L_{\mathrm{intrinsic}} = \left(\frac{10^{C}}{V_{\mathrm{norm}}}\right)^{1/\alpha}.
\end{equation}
Qualitatively, we simply use the $V_{\mathrm{norm}}$ for a quasar 
to read off 
an expected value of $L_{\mathrm{meas}}$ from the linear 
$\log(\overline{V}_{\mathrm{norm}})-\log(L_{\mathrm{meas}})$ plot.  
This expected $L_{\mathrm{meas}}$, 
as it is the prediction calculated using the luminosities of the entire 
quasar sample, is an estimate of $\overline{\mu} L_{\mathrm{intrinsic}}$, 
where $L_{\mathrm{intrinsic}}$ is the intrinsic luminosity of that quasar.  
Comparing this value 
to a direct measurement of the luminosity of the quasar $L_{\mathrm{meas}}$, 
we can measure any shift in the luminosity, i.e. magnification or 
demagnification:
\begin{equation}
\frac{\mu_{\mathrm{true}}}{\overline{\mu}} \equiv \frac{L_{\mathrm{meas}}}{\overline{\mu} L_{\mathrm{intrinsic}}} = L_{\mathrm{meas}} \left(\frac{V_{\mathrm{norm}}}{10^{C}}\right)^{1/\alpha}.
\label{mu_equation}
\end{equation}
We define $\mu' \equiv \mu_{\mathrm{true}}/\overline{\mu}$, the relative 
magnification with respect to the typical value for the whole sample.  
For a large set of quasars taken from a broad sky area, such as the 
data used in this work, we expect $\overline{\mu}$ to be close to unity.  
In this way, each variability measurement $V$, calculated from a pair of magnitude 
measurements $\Delta m$, yields an estimate of the quasar's relative magnification 
$\mu'$.


\section{Palomar-QUEST RG-610 Data Set}

\begin{figure}
\begin{center}
\includegraphics[width=85mm]{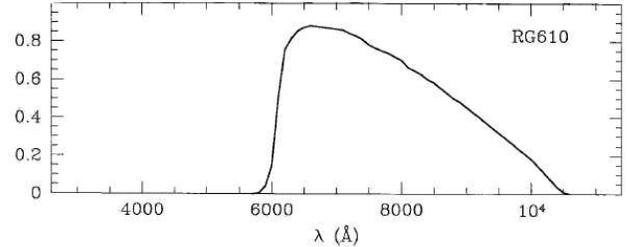}
\end{center}
\caption{RG-610 filter transmission, multiplied by CCD response.}
\label{rg610}
\end{figure}

We measure the lensing magnification of quasars using the 
Palomar-QUEST Variability Survey.  
The survey used the QUESTII large area camera 
(\citealt{camera_paper}), which comprises 112 CCDs with a pixel scale of 
0.878''/pixel, with an overall field of view of 9.6 square degrees.  
The camera was mounted on the 48'' Samuel Oschin Schmidt telescope at 
the Palomar Observatory, during which it aquired two main data sets.  First 
completed was a 3.5 year multi-color survey covering 15,000 square 
degrees roughly four times in each of seven optical filters 
(Johnson UBRI and SDSS $riz$), with 
a depth of roughly mag 21 in the SDSS $r$ filter.  These data have been 
used to study the variability of AGNs (\citealt{bauer09a}, 
\citealt{bauer09b}), to search for highly variable objects in both the 
archival data (\citealt{bauer09c}) and real-time data (\citealt{george}), 
and to study brown dwarfs (\citealt{cathy}, \citealt{cathy2}), 
among other phenomena.  The 
multi-color survey was executed concurrently with a 5 year, 30,000 
square degree survey which used a single broad, red, RG-610 filter, and 
reached a depth of about mag 19.5 in SDSS $r$ in each 60 second exposure 
(under dark sky conditions).  
The bandpass for the RG-610 filter (filter transmission times the QUEST 
CCD quantum efficiency) is shown in figure \ref{rg610}.  
The single-filter survey has been used as the discovery data for the 
Nearby Supernova Factory (\citealt{snfactory}), as well as for searches 
for dwarf planets and other solar system bodies (\citealt{planet}).  

This work uses the single-filter Palomar-QUEST data, and constitutes the 
first results from these data which rely on precision photometry.  For 
this reason, we describe the DeepSky image reduction pipeline and 
the photometric calibration techniques 
which are used to generate flux measurements.

\subsection{The DeepSky Image Reduction Pipeline}

The Palomar-QUEST RG-610 data were taken by a variety of groups with
different strategies, under different observing conditions, and with no
uniform calibration procedure; daily operations in each of the groups
emphasized simply discovering transients, rather than more precise
photometry.  Industry-standard reduction frames such as dark frames,
dome flats, twilight flats, and fringe frames,
were taken rarely or not at all.  The DeepSky project has re-reduced these
data using a standard pipeline designed to improve the precision of
differential photometry, which we briefly describe below.

The DeepSky reduction suite treats each of the 112 CCDs in the QUESTII
camera as a separate detector, attempting only to produce a uniform response
on each frame rather than trying to flatten the detector response across
the entire mosaic.  Data frames for each stage of the reduction
(dark, flat, etc.) were pooled for each month, and sometimes from several
months in succession depending on what was available, to achieve the high
statistics necessary to reject bad pixels and regions of the camera with
unstable performance.  The algorithms were cross-validated on subsamples of
the data for months with many reduction frames, suggesting that the loss
of performance is minimal when building reduction frames monthly rather than
more frequently (weekly or daily).  Each month about 100 calibration images
from each chip were used to create 'superframes' for the reduction.

Bias levels were first subtracted from each $600 \times 2400$ pixel 
image using a
$40 \times 2400$ overscan region.  Dark frames of different exposure times
(10, 60, 100, and 240~sec) were combined, with the dark current $D_i$ at
each pixel $i$ being fit to a model linear in the exposure time $t$: 
$D_i = a_i + b_i t$, 
where the 'instantaneous' portion corresponds to dark current which
accumulates during readout.  Iterations of the fitting procedure rejected 
outlier
values in determining the model parameters $a_i$ and $b_i$.  Pixels for which
no solution converged, or for which the ``1 $\sigma$'' interquantile width
(median minus 16th percentile) of the distribution of residuals from the
fit was more than three times wider than the 
median residual value for the image, 
were masked as bad pixels.  All images were then corrected according
to their exposure time using the superdark $D_i$.

In general neither twilight nor dome flats were regularly available,
and images taken under dark-sky conditions were often contaminated with
strong fringes.  We therefore chose to construct flat fields using science
images taken both during twilight and in moonlight.  
The calculations of \cite{ks91}, appropriately
adapted to Mt. Palomar, suggests that the loss of flatness due to the
non-uniformity of moonlight should be less than 2\% over the extent of each
chip (0.5 degrees) for images taken more than 20 degrees from the moon,
and we required this of images used to build the flat field.  We also required
that the sky level from scattered moonlight in these images be at least five
times that typical of dark sky conditions.  Images passing both of these cuts
were normalized to a mean sky brightness of 1.0, and combined by taking the
median at each pixel location to produce a superflat for the month.
Pixels with unusually large fluctuations relative to other pixels were masked
as bad, as for superdarks above.  All images were then flattened using these
superflats.

Superfringes were constructed in an analogous manner to the superflats
using images taken under dark-sky conditions.  The baseline fringe pattern
was assumed to be a time-invariant characteristic of each chip over the month.
The fringe amplitude in each science image was determined through
cross-correlation with the superfringe, and the fringes in the image removed
by subtracting the superfringe scaled by this amplitude.
Changing sky conditions can of course produce time variations in the fringes,
particularly in such a wide-band filter.  We accept this as a systematic
error in the photometry, with a contribution roughly on the same order as
the sky noise under dark sky conditions.

Object detection and aperture photometry were performed on the detrended 
(bias-subtracted, dark-subtracted, flattened and de-fringed) images using 
\texttt{SExtractor} (\citealt{sextractor}).  
Astrometry was performed on each frame using 
the \texttt{astrometry.net} suite (\citealt{astronet}).  
Magnitudes were measured for 
each object using ten aperture diameters ranging from 1 pixel to 16 pixels; 
 this work uses the 3-pixel diameter aperture measurements 
as the primary flux measurement.  
Aperture corrections were calculated as the clipped median of the 12-pixel 
diameter aperture flux divided by the 3-pixel aperture flux.  This correction 
was calculated using all good-quality measurements on a frame, and calculated 
separately for each frame.

\subsection{Photometric Calibration}

A number of cuts are made on the quality of the data.  First, 
\texttt{SExtractor} 
flags are checked for objects that are blended or saturated, close 
to image boundaries, or for which the measurement failed;  these 
flags eliminate about 3\% of the data.  
A further saturation check is made by comparing object fluxes to 
saturation levels determined for each chip, which are more accurate than 
the single baseline saturation level checked by \texttt{SExtractor}.  
Very few objects are eliminated by this cut.  
Variations in the background sky level on the small spatial scale of 
5$\times$5 pixels 
are examined, and a frame is discarded if they are typically large 
enough to contribute a 3\% error to the photometry of an object with 
magnitude 18;  this removes about 3\% of the images.  Frames are also 
rejected for which the median full width at half maximum (FWHM) 
exceeds 5 pixels, eliminating about 6\% of the data.  
To ensure that closely spaced detections do not contaminate the aperture 
photometry, an object is rejected if it has a neighbor within four arc 
seconds ($\sim$3.5 pixels);  most object affected by this cut are 
already discarded for being blended in \texttt{SExtractor}.

Spatial fluctuations of the moon/twilight flat fields 
and variability in night sky lines during observations made it 
difficult to completely remove the fringes from the data.  This
results in a $\sim 2\%$ flux systematic for bright objects, which is 
added in quadrature to the statistical errors.

A standard method of calibration is simply to determine a multiplicative 
constant zero point which corrects differences in average response 
(both intrinsic 
and weather-related) between two images.  
We perform such a frame-based calibration as a first step.  All overlapping 
images are determined, and the one in which objects have the brightest 
measurements is assigned as the standard.  A single zero point 
is determined for each frame in order to bring it to the 
level of the standard.  This constant is determined using objects 
of all magnitudes.  The correction is also calculated separately 
for bins of different magnitudes:  five bins between SDSS r magnitudes 
of roughly 15 and 20.  On average, the difference in the mean correction 
between the brightest bin 
and a fainter one ranges from about 3\% to 6\%.  If a larger variation 
than this is seen, the frame is assumed to be non-linear and is not 
used.  This eliminates about 15\% of the data;  the non-linearity is 
partially due to intrinsic properties of the CCDs, but primarily due to 
the insufficient quality of the fringe and flat field calibration data.  
Poor frames are also identified by comparing their measurements to the 
mean measurements of the objects they hold.  If 15\% of the flux 
measurements on a frame disagree with their objects' means by 
more than 3 times the measurement errors, 
then the frame is discarded.  This removes about 8\% of the data, 
for which the systematic effects are catastrophic.

The systematic residuals in the data make a frame-based calibration 
insufficient for accurate photometry;  
while a systematic term describes the typical uncertainties, the error 
distribution is not Gaussian and therefore this procedure generates 
many photometric outliers.  Because the spatial scale of the systematic 
variations is much larger than the size of each object, we can refine 
the frame-based calibration using an object-based one.  This technique 
involves generating a zero point for one object at a 
time by considering only its close neighbors.  A large scale systematic 
variation far away from an object will affect a frame-based calibration; 
an object-based one will be less sensitive to such features.  
We therefore implement, as a second step, an 
object-based calibration. We use the average of at least 10 neighbors 
within 150 arc seconds to calculate the correction between an object's 
measurements on two scans.  About 8\% of the detections do not have enough 
neighbors to satisfy this requirement, and are therefore disregarded.

The frame calibration is calculated using only 
objects detected by the Sloan Digital Sky Survey (SDSS; \citealt{sdss7}) 
and classified by them 
as pointlike.  This precaution eliminates the possibility of our 
inadvertently using data artifacts, or galaxies which may exceed 
the aperture size, as calibration objects.  
For the purposes of this work, the object calibration is only 
performed on quasars which have been spectroscopically identified 
by SDSS.  
The reference objects used in the object calibration are limited 
to pointlike SDSS detections.

The strict cuts implemented in the photometric calibration throw 
away roughly 35\% of the measurements.  The systematic error is 
typically 2\%, with about 3\% of measurements disagreeing with the 
mean flux of the object by over 3 times their measurement errors.  
From previous variability 
studies we expect $< \sim$1\% of measurements to be truly variable 
(\citealt{huber06}; \citealt{sesar07});  therefore, we infer that our 
photometry contains 
residual calibration errors in $\sim$2\% of the measurements.  
Because in this work we deal only with the ensemble average over hundreds 
of measurements, and we expect the systematic error to be uncorrelated 
across different observations, 
we do not expect the residual errors to affect our results.  
Nonetheless, as a precaution against these errors we make additional 
cuts against variability behavior uncharacteristic of quasars;  this 
is described further in section \ref{datacuts_section}.

\section{Results}

We study a sample of 3573 spectroscopically identified quasars, 
spanning roughly 5,000 square degrees, from the SDSS data release 7.  
For each quasar we calculate its luminosity at 2500 \AA\ by multiplying 
a redshifted composite quasar spectrum taken from \cite{compositespectrum} 
with SDSS filter curves.  
We use the SDSS $z$ band magnitude, corrected for galactic extinction 
using the maps from \cite{schlegel}, to normalize the composite spectrum's 
amplitude.  We choose the $z$ band as it is least extincted by 
dust along the line of sight.  
We then take the normalized flux at 2500 \AA\, and convert 
it to luminosity using the object's spectroscopic redshift and 
the cosmological parameters, which we assume to be $\Omega = 1, 
\Omega_{\Lambda} = 0.7, \Omega_{M} = 0.3, H_{0} = 0.71 \frac{km}{s \cdot Mpc}.$  
We also estimate the black hole mass for each quasar from the $M-\sigma$ 
relation, using the widths of the $\mathrm{H_{\beta}}$ and $\mathrm{Mg_{II}}$ 
emission lines 
as in \cite{salviander07}, and calculating the radius of the BLR as 
described in \cite{bentz09}.

In many cases during the Palomar-QUEST Survey, the same sky area was 
observed twice in the same night.  Because quasars are not variable 
on such short timescales to the precision of our measurements, we 
average all intranight measurements to yield one flux value per night.  

\subsection{Data Cuts \label{datacuts_section}}

As described above, we have made restrictions on the data quality in order 
to produce reliable photometry.  Here we describe the criteria we use 
to select our quasar sample so that we have a well-measured, homogeneous 
dataset.

While the majority of quasar variability is thought to be due to 
accretion disk instabilities, some AGN show blazar characteristics, 
with considerable flux 
and variability from a jet component.  Jet-based variability 
is much more dramatic than that seen in typical quasars, and because 
it is powered by shock dissipation and enhanced by relativistic beaming, 
it is not 
expected to follow the variability-luminosity relation seen in most quasars.  
It is therefore important to restrict the quasar list to those 
displaying variability that is characteristic of quasars.

We select quasars from SDSS that have the following properties:

\begin{enumerate}

\item spectroscopic identification in the SDSS, with a match to the 
spectral cross-correlation template number 29 (QSO) with confidence 
of at least 0.95

\item no detection in the FIRST \citep{becker95} or RASS 
\citep{anderson07} catalogs

\item redshift between 0.4 and 2.2

\item estimated black hole mass between $10^{8} \mathrm{M_{\odot}}$ and $7.5 \times 10^{9} \mathrm{M_{\odot}}$

\item luminosity between $10^{30.85} \frac{\mathrm{erg}}{\mathrm{s} \cdot \mathrm{Hz}}$ and $10^{31.5} \frac{\mathrm{erg}}{\mathrm{s} \cdot \mathrm{Hz}}$.

\end{enumerate}

Criteria 1 and 2 are meant to select radio-quiet AGN whose optical fluxes do 
not show significant contributions from a jet.  Numbers 3-5 serve in part to 
crop the tails of the parameter distributions, and are the edges of 
the binning scheme that we use when normalizing the data as described 
below.  The bin limits are stated again in table \ref{bin_limits}.  The 
low redshift cut is chosen to eliminate objects that may appear 
spatially extended in the data, and therefore be poorly measured by 
the Palomar-QUEST photometry.  The high redshift cut ensures that 
the quasar sample is selected homogeneously using UV-excess techniques.  
The low luminosity cut is chosen because for objects fainter than this 
level we see a flattening in the variability-luminosity relation, 
implying that these less luminous quasars vary to fluxes below our 
detection limit, causing us to not observe their full range of variability.  

These criteria yield a sample of 4845 quasars that have Palomar-QUEST repeated 
measurements.

Furthermore, the quasars' Palomar-QUEST measurements must satisfy:

\begin{enumerate}

\item $V \leq 0.1$ where $V$ is defined as in 
equation \ref{v_equation} and is calculated using all measurement pairs 
with rest-frame time lag less than 10 days

\item $V \leq 0.5$ where $V$ is defined as in 
equation \ref{v_equation} and is calculated using all measurement pairs 
with rest-frame time lag greater than 100 days.

\end{enumerate}

These cut values 
are motivated by the variability amplitudes of quasars and blazars 
measured in \cite{bauer09b}.  Because the 
variability amplitude is unchanged for lensed and unlensed objects, 
this is an unbiased cut with respect to lensing analyses.  
This step eliminates 967 objects, leaving 3878 quasars.  We note 
that this cut eliminates a much larger fraction of the AGN than is 
expected from the relative numbers of known blazars and quasars.  
However, the cut is not only sensitive to jet-based variability but also 
calibration errors, and is a strict but important criterion to ensure 
that poor quality measurements do not overwhelm the lensing signal.  
These 3878 quasars have, in total, 230,674 Palomar-QUEST measurement 
pairs.

We only use measurement pairs for which:

\begin{enumerate}

\item the rest frame time lag between measurements is shorter than 400 days

\item the measurements show significant ($> 1 \sigma$) variability

\item the variability amplitude is less than 1 magnitude

\item no more than 40 measurement pairs from each quasar are included.

\end{enumerate}

Cut 1 eliminates a long but low-level tail out to larger time lags, 
and throws away 19,523 pairs.  Cut 2 is necessary to avoid the 
calculation of imaginary $V$ values for individual 
measurement pairs and is very restrictive because the Palomar-QUEST 
systematic error is of the order of typical quasar variability amplitudes over 
rest frame timescales of weeks.  Still, as will be seen in figure 
\ref{v_vs_l2500}, the data after this cut continue to follow a 
variability-luminosity relation described by equation 
\ref{v_vs_l_equation}, and therefore can be used to measure 
magnification using equation \ref{mu_equation}.  We also note that 
removing the lowest signal-to-noise $V$ values reduces error due to 
imprecise determination of the survey's systematic uncertainties, 
given the dependence of $V$ on the measurement uncertainty as stated 
in equation \ref{v_equation}.  Because the systematic uncertainties 
stem in part from imperfect flat field and fringe calibration images, 
they can be time and position dependent and therefore difficult to 
determine precisely for each measurement.  
This cut eliminates 134,745 pairs.  

Cut 3 is a final check against outliers 
that are uncharacteristic of quasar behavior, and only throws away 17 
measurement pairs.  Cut 4 serves to keep a minority of 
exceptionally well-measured 
quasars from dominating the dataset, since the number of measurement 
pairs rises as the number of measurements squared.  This cut 
removes 14,671 pairs.

Normalization of the quasars, described in section \ref{mu_meas_section}, 
is only done if there are sufficient quasars with similar 
properties so that we can determine accurate zero points.  
We require 50 measurement pairs per $\tau$ $M$ $z$ $L$ 
bin, which removes 4359 measurement pairs.  

We are left with 57,359 useable measurement pairs, from 3573 quasars.  
We note that many of our cuts are due to properties of the 
data such as statistics and measurement errors, rather than properties of 
the quasars.  Cuts on future quasar data sets may not need to be 
so restrictive.


\subsection{Variability-Luminosity Relation}

\begin{figure}
\begin{center}
\includegraphics[width=85mm]{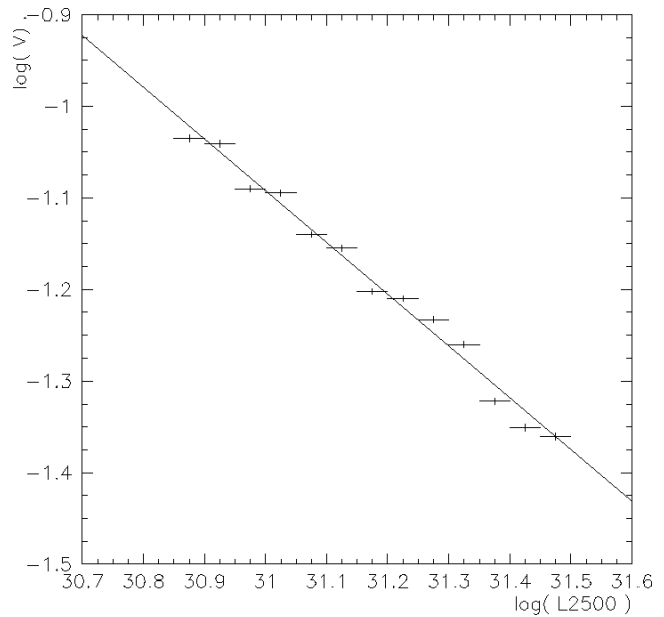}
\end{center}
\caption{Log Variability vs. Log Luminosity at 2500\AA, after normalizations, determined using 57,359 measurement pairs.  The best-fit line is shown, with slope $-0.565 \pm 0.007$.}
\label{v_vs_l2500}
\end{figure}

After normalizing the data as described in section \ref{mu_meas_section}, 
we plot the normalized variability vs. luminosity relation, 
which is shown in figure \ref{v_vs_l2500}.  
The $y$ axis error bars on the points are errors on the mean calculated 
from the spread of the measurements in each bin.  
The bins in the figure are smaller than those used in the normalization, 
in order to show the relation in more detail.  
The trend is indeed well-described by a 
power law, with index $\alpha = 0.565 \pm 0.007$.  
The best fit line is shown on the plot.  This slope is steeper than 
that found in other works (e.g., \citealt{vandenberk04}, \citealt{bauer09a});  
however, it is 
not directly comparable since we only consider measurements for which 
we observe significant variability.

The precision of our magnification measurements relies on a narrow 
spread in the variability-luminosity relation.  The residuals 
in log($V$) around the best-fit line are distributed normally, with 
a standard deviation $\sigma = 0.22$ that is constant across our range 
of luminosity.  These residuals are larger than the 
typical measurement errors, and therefore reflect an inherent 
spread in the quasar variability properties.  
%
%
There are several possible 
sources of the enhanced variance.  

The measurement error on the luminosity is small, and is based simply on 
the uncertainties in the SDSS broadband magnitudes.  This error estimate 
ignores the fact that the quasars vary by more than these uncertainties;  
the true mean luminosity could be tens of percent different from 
what was measured when the SDSS made its photometric observation.  Since 
the spectroscopic quasar sample is much brighter than the SDSS photometric 
detection limit, there should be no Malmquist bias in the quasar sample 
and the error on the measured luminosity should be unbiased.  Therefore, 
when discussing the variance of the relation, we assume that the luminosity 
is correct and simply absorb its error into the $y$ axis spread in values.  
In future studies this uncertainty can be reduced in cases where 
there are more photometric measurements per quasar, which can be averaged 
to produce a more accurate typical luminosity.  (This is not possible 
with the Palomar-QUEST data, as it is relatively calibrated rather than 
absolutely.)  Or, if many measurements 
are available, a lightcurve may be fit to each quasar's data to 
determine the object's baseline luminosity most accurately 
\citep[see, e.g.,][]{macleod10}.

If the spread in the variability-luminosity relation is due to an 
effect with zero bias, then all quasars will 
intrinsically lie on the mean relation and increased statistics for a given 
object will move it closer to the mean trend.  If quasars in the 
sample have heterogeneous variability properties, an object may lie 
intrinsically off the variability-luminosity relation and more statistics 
will only reveal this property more clearly.  Some insight into this 
issue is given in \cite{macleod08}, which explores the implications 
of using many, as opposed to simply two, 
measurements of each quasar when studying the objects' variability 
properties.  They note that using only two measurements yields 
ensemble variability results similar to those determined using many 
epochs from the SDSS stripe 82 region.  However, they also point out 
that there is an intrinsic spread in the quasars' variability behavior 
that is not currently understood.  This intrinsic difference between 
individual objects surely contributes to the scatter observed in the 
variability-luminosity relation, and is our motivation for including 
no more than 40 measurement pairs from each object in this analysis.  
More insight into the physical processes driving the different 
variability properties of individual quasars would help in 
understanding and reducing the systematic errors in the 
lensing magnification measurement.

Assuming statistically identical quasar lightcurves with power law 
spectra, \cite{bauer09a} used Monte Carlo simulations to show 
that using few measurements per quasar to calculate $V$ versus $\tau$ 
gives a slope that is unbiased, although windowing effects can make 
the trend non-linear at very short and long time lags (which are excluded 
from the data set in this work).  Therefore we expect no significant 
bias to be introduced in this analysis due to the behavior of $V$ in 
the presence of sparse data sampling.

%
%

\subsection{Magnification Measurements}

\begin{figure}
\begin{center}
\includegraphics[width=85mm]{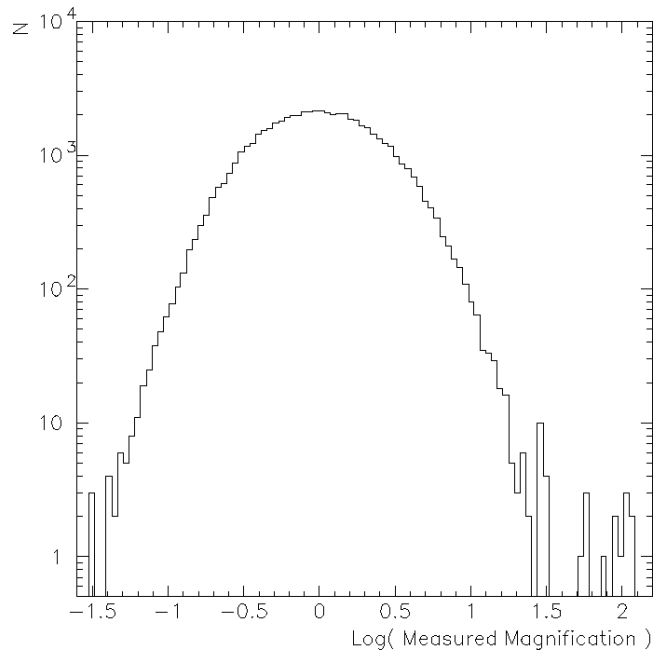}
\end{center}
\caption{Logarithm of the measured quasar magnifications, determined using 57,359 measurement pairs.}
\label{measured_mu}
\end{figure}

Now that we have calibrated the quasar measurements and confirmed the 
variability-luminosity relation, we calculate the magnification 
$\mu' \equiv \mu_{\mathrm{true}}/\overline{\mu}$ as defined in equation 
\ref{mu_equation}.  The distribution of $\log(\mu')$, with one entry per 
quasar measurement pair, is shown in figure \ref{measured_mu}.  
The distribution is wide and close to Gaussian. 
This is due to the distribution of the residuals of the $V-L$ relation, 
which has a similar shape.  Because the scatter in the $V-L$ relation 
is large, the width of the $\log(\mu')$ distribution is much wider than 
the range of magnifications expected in the data.  
Just as the mean of many variability measurements produces a precise 
variability-luminosity relation, the mean of many magnification 
measurements produces a meaningful result, with an error on the mean 
much smaller than the width of the distribution.

In order to obtain mean magnifications, we must bin the measurements 
in a physical quantity 
which we believe to be related to the magnification.  Here we choose 
to bin the measurements in terms of their distance perpendicular to 
the line of sight 
(scaled as distance/$R_{200}$) from the nearest member of the MaxBCG 
galaxy cluster catalog.

\subsection{The Expected Signal} \label{expected_signal_section}

The strongest magnification of quasars is expected to be due to clusters 
of galaxies along the line of sight.  To estimate the level of 
magnification that we expect quasars to experience, we can calculate 
the magnification due to the members of known cluster catalogs.  
For this purpose we use the MaxBCG catalog, which contains 
13,823 clusters and covers about 7,500 square degrees.  
The MaxBCG catalog 
is estimated to be 90\% pure and 85\% complete for clusters between 
redshifts 0.1 and 0.3 and with masses greater than $10^{14}M_{\odot}$ 
\citep{maxbcg}.  
By adopting a model for each cluster, we 
can calculate the expected magnification of each quasar due to the galaxy 
clusters in the catalog.  

We assign a Navarro-Frenk-White (NFW) profile (\citealt{nfw}) to each cluster:
\begin{equation}
\rho(r) = \frac{\delta_{c} \rho_{\mathrm{crit}}}{(r/r_{s})[1+(r/r_{s})]^{2}}
\label{nfw_equ}
\end{equation}
where $c$ is the concentration parameter of the profile, 
$\delta_{c} = \frac{200}{3} \frac{c^{3}}{\mathrm{ln}(1+c) - c/(1+c)}$, 
$r_{s} = R_{200}/c$, 
and $R_{200}$ is the radius inside which the density is 200 
$\times \rho_{\mathrm{crit}}$, the critical density of the universe at the 
redshift of the cluster.   The total mass inside $R_{200}$ is $M_{200}$.  
We assume the cluster mass-richness relation used by \cite{rozo09}: 

\begin{equation}
M_{200}(N_{200}) = \frac{1}{1.022} e^{0.48} \times \left(\frac{N_{200}}{20} \right)^{1.13}
\label{rozo_equ}
\end{equation}
to estimate $M_{200}$ given the richness $N_{200}$ stated in the cluster 
catalog.  
$M_{200}$ and $c$ have been seen to correlate with each other;  we 
use the mass-concentration relation from \cite{duffy08}:

\begin{equation}
c(M_{200},z) = 5.71 \times \left(\frac{M_{200}}{2.e12 \times h^{-1}}\right)^{-0.084} \times ( 1 + z )^{-0.47} 
\end{equation} 
 to determine $c$ given $M_{200}$ and the cluster redshift $z$.  When 
calculating the magnifications, we truncate each profile at 
$3 \times R_{200}$.  If a quasar is close to more than one cluster, 
we multiply the magnifications due to each nearby cluster to produce a 
magnification estimate that includes contributions from all known 
clusters.  Multiple magnifications are a minor effect;  only 6.5\% 
of the quasars have expected magnifications of $>1\%$ from more than 
one cluster.

\begin{figure}
\begin{center}
\includegraphics[width=85mm]{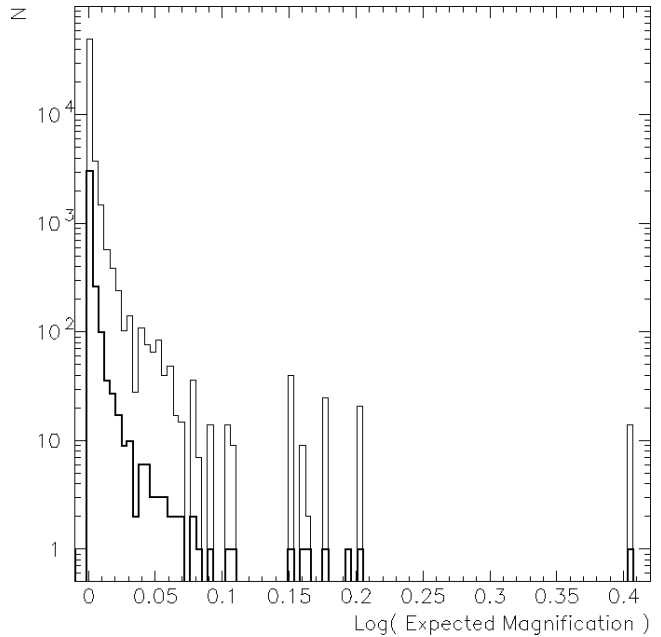}
\end{center}
\caption{Logarithm of the lensing magnification of quasars assuming NFW profiles for MaxBCG galaxy clusters.  The thick-lined histogram includes one entry for each of the 3573 quasars which overlap the MaxBCG area.  The thin-lined histogram shows the same calculated magnifications, but includes one entry for each of the 57,359 quasar measurement pairs and is therefore comparable to figure \ref{measured_mu}.}
\label{expected_mu}
\end{figure}

A histogram of the logarithm of the predicted magnifications for the 
quasar sample is shown in figure \ref{expected_mu}.  
We predict one magnification per 
quasar, but we measure one $\mu'$ per pair of magnitude measurements.  
The thick-lined histogram in figure \ref{expected_mu} shows the distribution of 
magnification predictions, with one entry per quasar and therefore 3,573 
entries.  The thin-lined 
histogram includes one entry for each magnitude measurement pair:  if 
a quasar has $N$ measurements, then that quasar's predicted magnification 
will have $N(N-1)/2$ entries of identical value.  The thin histogram 
therefore has 57,359 entries and is comparable to figure \ref{measured_mu}.  
The vast majority of the quasars are not 
significantly magnified by the clusters;  however, there is a small 
tail of highly magnified objects up to $\mu \sim 2$.  
The striking difference in the shapes of the expected magnification distribution 
(figure \ref{expected_mu}) and the measured distribution (figure \ref{measured_mu}) 
is due to the large scatter in the individual measured values, reflecting the scatter in the 
$V-L$ relation.  These figures visually demonstrate the need to average many 
magnification measurements in order to obtain a result precise enough to compare 
to theory/models.

\begin{figure}
\begin{center}
\includegraphics[width=85mm]{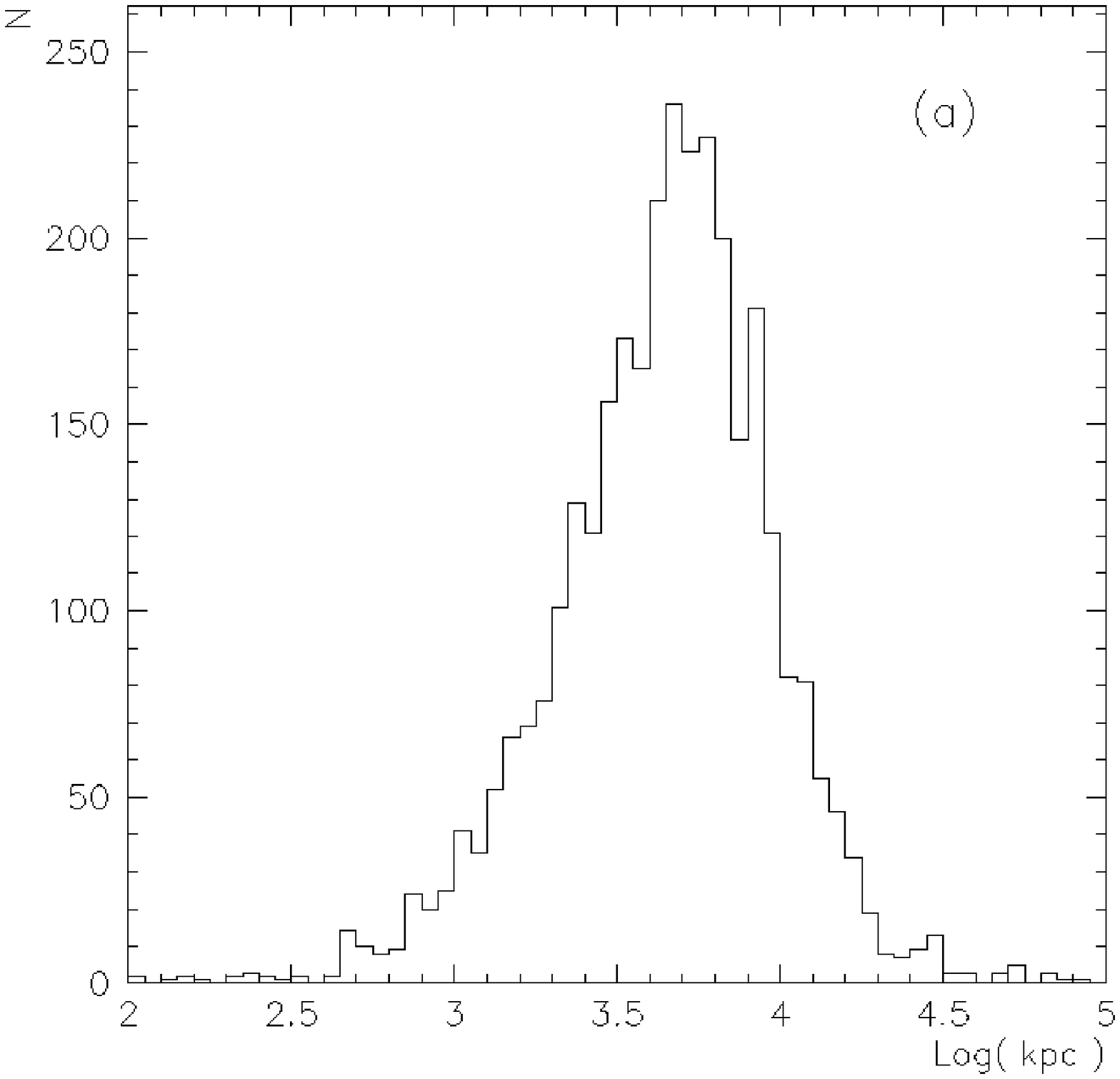}
\includegraphics[width=85mm]{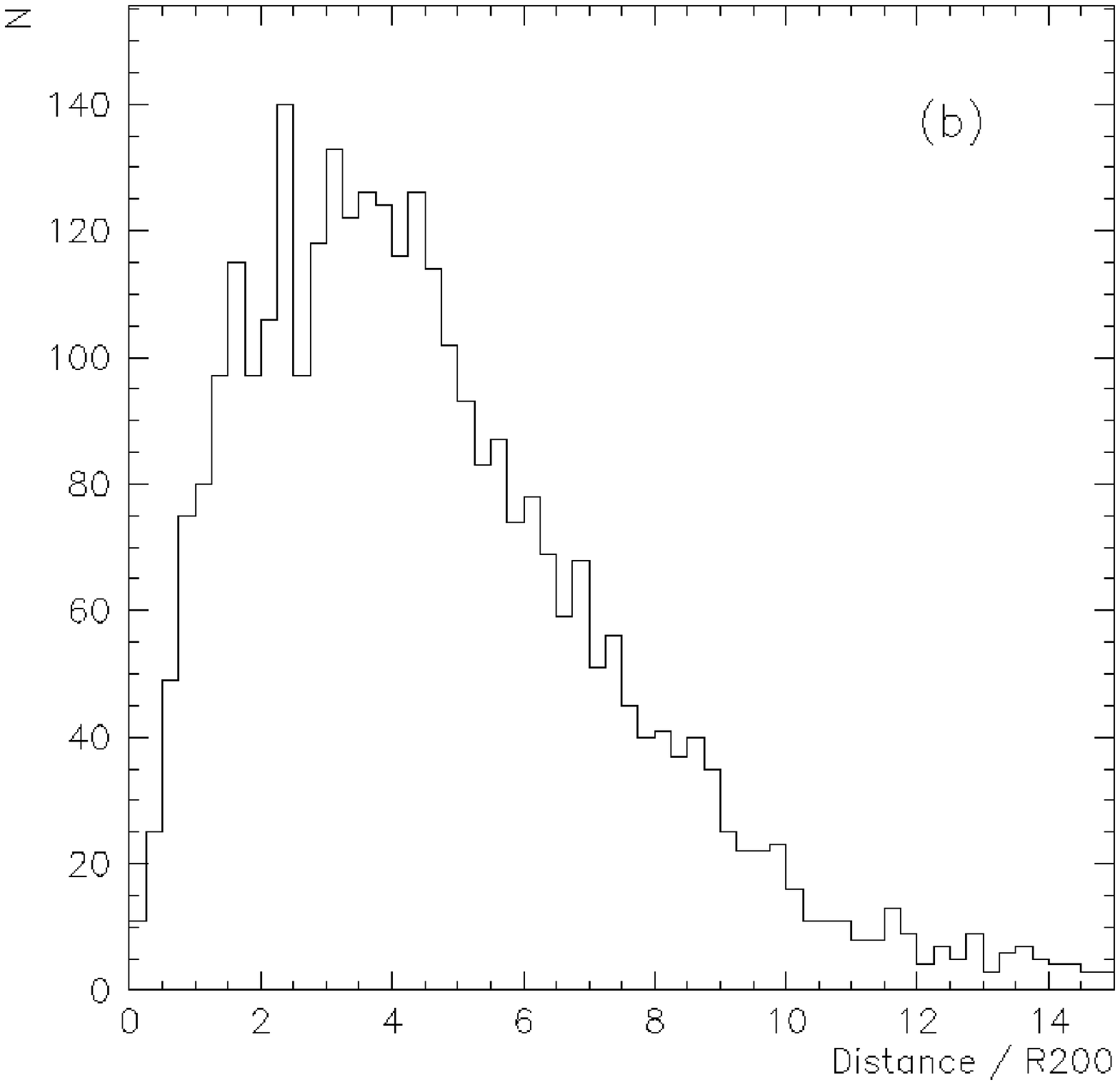}
\end{center}
\caption{Distance, transverse to the line of sight at the redshift of the cluster, between the quasars and the clusters which are closest on the plane of the sky.  (a) in units of log kiloparsecs at the redshift of the cluster.  (b) in units of $R_{200}$ of the closest cluster.}
\label{qso_cluster_dist}
\end{figure}

The transverse distance between each 
quasar and the MaxBCG cluster closest on the plane of the sky is shown 
in figure \ref{qso_cluster_dist};  panel (a) shows the distance in the 
logarithm of kiloparsecs at the redshift of the closest cluster, while panel 
(b) shows the distance as a fraction of $R_{200}$ of the closest cluster.  

\subsection{Magnification profile of the MaxBCG Clusters}

For each quasar, we calculate the magnification we expect due to 
clusters in the MaxBCG catalog, as described in section 
\ref{expected_signal_section}.  We then group the quasars in 
logarithmically spaced bins given by the quasar's distance from 
the closest cluster, in units of $R_{200}$ of that cluster.  
The errors on the measurement are determined through bootstrap 
resampling.  In particular, for each bin 1000 mean magnifications 
are calculated using random sampling with replacement such that the 
randomized set includes the same number of measurements as the 
original bin.  The errors are taken to be the $1 \sigma$ 
(15.9 and 84.1 percentile) values of the 1000 mean magnifications.  
Because the errors on the magnifications are in part caused by 
systematic uncertainties that are not currently understood, we prefer 
to derive the errors empirically in this way using the properties of 
the distribution itself.

\begin{figure}
\begin{center}
\includegraphics[width=85mm]{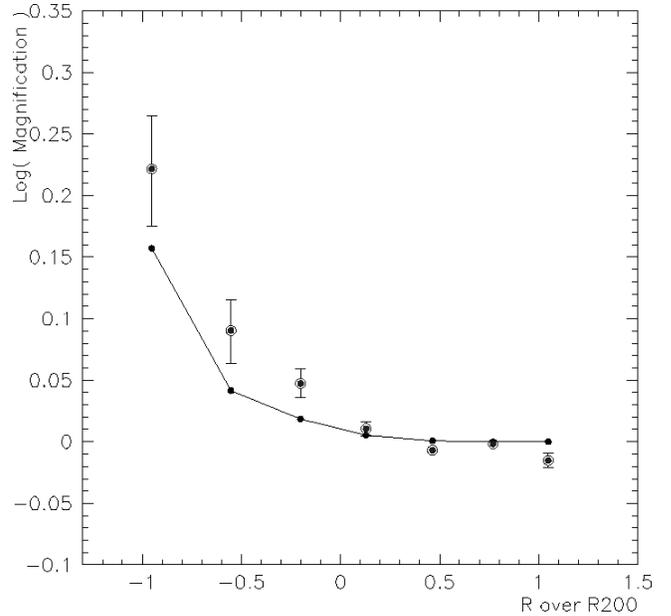}
\end{center}
\caption{Measured $\mu'$ (points with error bars) and expected $\mu$ (points connected by lines), versus distance from the nearest cluster in units of $R_{200}$, assuming NFW profiles truncated at $3 \times R_{200}$.}
\label{results_rozo}
\end{figure}

Figure \ref{results_rozo} 
shows the mean measured $\mu'$ (points with error bars) and the expected 
magnification (points without errors, connected by lines), 
assuming NFW haloes for the MaxBCG catalog's clusters as described above.  
While the magnification is significant and does fall with radius, 
we measure a systematically different magnification profile from that 
expected.  The measured values compare to the expected magnifications 
with reduced $\chi^{2} = 1.84$.

\begin{figure}
\begin{center}
\includegraphics[width=85mm]{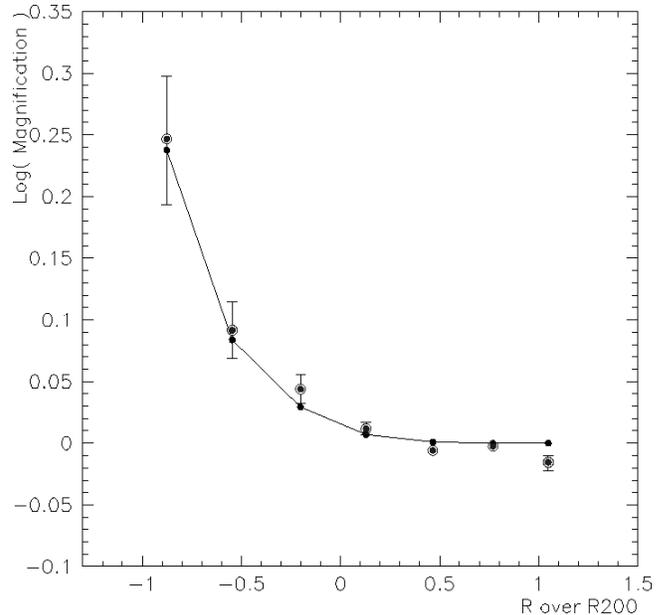}
\end{center}
\caption{Measured $\mu'$ (points with error bars) and expected $\mu$ (points connected by lines), versus distance from the nearest cluster in units of $R_{200}$, assuming NFW profiles truncated at $3 \times R_{200}$ and with centers offset as prescribed in \cite{johnston08}.}
\label{results_rozo_offset}
\end{figure}

The MaxBCG catalog's cluster positions are given as the location of the 
brightest cluster galaxy (BCG).  This position does not always coincide 
with the center of the dark matter halo.  Errors in the cluster 
central coordinates cause the average radial profile of the MaxBCG 
clusters (as measured from the reported centers) to differ from 
NFW.  \cite{johnston08} studied this effect using weak lensing shear 
measurements and N-body simulations, and found a Gaussian distribution 
of positional offsets between the true and measured centers, with a 
richness-dependent probability that the cluster is correctly centered.  
To determine the effects of such offsets on our expected magnifications 
we implement the prescription of \cite{johnston08}, offsetting 
the center of a subset of the clusters in a random manner.  The results 
are shown in figure \ref{results_rozo_offset}, and provide an 
improved fit to the data, with reduced $\chi^{2} = 1.17$.  
The agreement between the data and predictions may be further improved by 
modelling the effects of large scale structure outside the redshift 
range of the MaxBCG catalog;  however, these effects are subdominant 
and beyond the scope of this paper.

\begin{figure}
\begin{center}
\includegraphics[width=85mm]{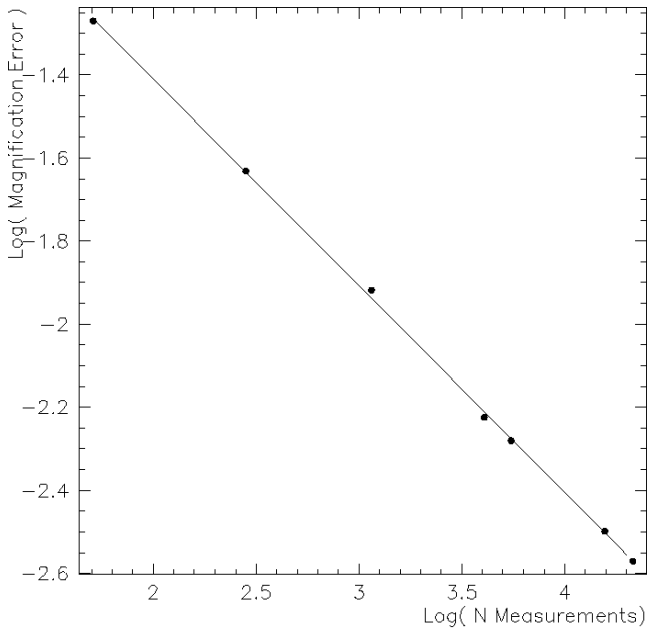}
\end{center}
\caption{Logarithm of the error on each bin versus the number of measurements in that bin, taken from the data in figure \ref{results_rozo_offset}, with a best-fit line superimposed.  The trend falls as $\sqrt{N}$, implying an error on $\mathrm{log}(\mu')$ of 0.04 for a magnification measurement using one pair of magnitude measurements. }
\label{errors_figure}
\end{figure}

Using the errors on the binned measurements determined through bootstrap 
resampling, 
we can estimate the error on each magnification measurement.  
Figure \ref{errors_figure} shows the logarithm of the error on each bin 
versus the number of measurements in that bin, taken from the 
data in figure \ref{results_rozo_offset}, with a best-fit line 
superimposed.  The trend falls as $\sqrt{N}$, and at the level of our 
most populated bin (with 21,482 measurement pairs) shows no sign of 
a systematic floor.  The best-fit line extrapolates to imply that 
each $\mathrm{log}(\mu')$ measurement, made using one magnitude measurement pair, 
has an error of roughly 0.4.  For a magnification of 1.1, this corresponds 
roughly to a signal to noise of 1 for each magnification measurement.

\section{Discussion and Conclusions}

We introduce a new technique to measure the effects of gravitational 
lensing by large scale structure.  Type I quasars on average follow a 
well-constrained 
variability-luminosity relation;  lensing magnification causes a 
significant shift in this relation.  We describe the magnification 
measurement procedure, and use it to determine the magnification of 
background quasars due to 
lensing by the galaxy clusters in the MaxBCG catalog.  

The measured magnification is significant, and consistent with the 
signal expected 
from the catalog.  In particular, the agreement between the 
data and expectations is improved when the 
miscentering of the clusters is taken into account, as described in 
\cite{johnston08}, compared to when the clusters are assumed to be NFW 
profiles centered exactly on their catalog positions.  
The lensing effects of the MaxBCG catalog have been measured very 
precisely using weak lensing shear, for example in \cite{sheldon09}.  
Although the magnification measurements presented here are not 
competitive with such results, they serve to show that quasar 
variability can be used to study lensing magnification, and are 
the first measurements using this new technique.

Using quasar variability to measure gravitational magnification is a very 
different approach from the common method of using galaxy shapes to 
measure gravitational shear.  Precision shape measurements of distant 
galaxies require very good quality data with faint flux limits and small 
point spread functions.  Furthermore, the complex intrinsic shapes of 
galaxies make the shear measurement nontrivial.  Measuring the variability 
amplitudes of quasars is much more simple, as these objects are often bright 
and they typically vary by amounts larger than the photometric errors of 
surveys such as Palomar-QUEST or SDSS.  We see the error 
on the measured magnification to be roughly 
given by 
\begin{equation}
\log(\sigma_{\log \mu}) = -0.5 \log(N_{\Delta m}) - 0.4 
\end{equation}
where $N_{\Delta m}$ is the number of $\Delta m$ measurements.  
This can be compared to the error 
on weak lensing measurements of the reduced shear $g$, which scale as 
$\sigma_{g} = \sigma_{\epsilon} \times (1 - |g|^{2}) / \sqrt{N_{\mathrm{gal}}} \approx 
\sigma_{\epsilon} / \sqrt{N_{\mathrm{gal}}}$ for moderate shears.  
The intrinsic scatter in the galaxy shapes is 
$\sigma_{\epsilon} \sim 0.35$ (\citealt{schneider04}), and $N_{\mathrm{gal}}$ is 
the number of measured galaxies.  
For magnification by a galaxy cluster with an NFW profile parameterized 
by $M_{200} = 10^{14}$ and concentration $c = 6$, a region with moderate 
magnification $\mu = 1.05$ would have reduced shear $g = 0.035$.  In this 
regime, 100 galaxies and 100 $\Delta m$ measurements would yield signal to noise 
$S/N_{\mu}$ = 10 and $S/N_{g}$ = 1.  
A region of higher magnification $\mu$ = 1.25 and $g$ = 0.083 would 
yield again $S/N_{\mu}$ = 10, with $S/N_{g}$ = 2.4.  
This lensing strength is at the upper end of the range for which 
weak lensing techniques have been rigorously tested, for example by the 
STEP2 project (\citealt{step2}).  Stronger distortions are not 
straightforward to measure, particularly using the common KSB 
method (\citealt{ksb}).  The variability-based quasar magnification 
analysis, on the other hand, is equally applicable to all lensing 
regimes.  

The fact that the magnification 
measurements each have higher signal to noise than the shear measurements 
is offset by the relative sparsity of quasars compared to inactive galaxies.  
In the current SDSS spectroscopic sample, which reaches depths of $r$ $\sim$ 
19.5, there are about 15 quasars per square degree.  In data down 
to $r$ $\sim$ 20.5 we expect roughly 40 quasars per square degree 
(\citealt{richards06}).  This depth will soon be available over large areas 
of sky.  For example, BOSS is expected to double the number of 
spectroscopic quasars in the SDSS Survey over the next several years, 
over a footprint of 10,000 square degrees\footnote{http://www.sdss3.org/}.  
Shear analyses are usually performed over much smaller areas of high 
quality data.  For example, results from the CFHT Wide Survey use 
50,000 galaxies per square degree, down to a magnitude of $i$ $\sim$ 24.5, 
with a total area of 34 square degrees (\citealt{fu08}).  
In order to reach the same signal to noise for moderate lensing strengths 
of $\mu = 1.05$ and $g = 0.035$, assuming 40 quasars and 50,000 galaxies 
per square degree, we would need fewer than 4 magnitude 
measurements on average for each 
quasar (assuming that the measurements give uncorrelated information 
about the magnification;  more work is required to investigate this 
systematic complication).  Assuming shallower data with 15 quasars 
per square degree, we would require 6 magnitude measurements of each 
quasar, on average, over just 34 square degrees.  Alternatively, 
the same signal to noise could be achieved with just 2 measurements per 
quasar over an area of 1100 square degrees.  
These levels of data are easily obtainable with surveys 
such as BOSS and Pan-STARRS, which will provide $>$10,000 square degrees 
with sufficient data quality for magnification measurements.  This is in 
sharp contrast with the $\sim 70$ 
square degrees which will be available from Pan-STARRS Medium Deep fields 
with the quality assumed here for shear measurements.  

This work introduces the technique of using variability to measure 
the lensing magnification of quasars, and measures the magnification of 
3573 quasars, with magnitudes down to SDSS r $\sim$ 19.5, due to lensing 
by known galaxy clusters along the line of sight.  Future large-area 
sky surveys will be able to apply this new technique to measure lensing 
magnification over larger volumes, and improve 
our understanding of large scale structure.

\section*{Acknowledgments}

We thank Peter Nugent and Janet Jacobsen for their work in developing 
and running the DeepSky pipeline.  
We acknowledge the support of the European DUEL Research Training Network, 
Transregional Collaborative Research Centre TRR 33, the Cluster of 
Excellence for Fundamental Physics, 
Spanish Science Ministry
AYA2009-13936, Consolider-Ingenio CSD2007-00060 and project
2009SGR1398 from Generalitat de Catalunya.  
The National Energy Research Scientific Computing Center,
which is supported by the Office of Science of the U.S.
Department of Energy under Contract No. DE-AC02-05CH11231,
has provided resources for this project by supporting staff,
providing computational resources and data storage.  
We thank the Office of Science of the Department of Energy
(grant DE-FG02-92ER40704) and the National Science Foundation
(grants AST-0407297, AST-0407448, AST-0407297) for support.  

\bibliography{bibliography}
\bibliographystyle{apj}

\end{document}